\documentclass[twocolumn,showpacs,preprintnumbers,amsmath,amssymb]{revtex4}
\usepackage{graphicx}% Include figure files
\usepackage{dcolumn}% Align table columns on decimal point
\usepackage{bm}% bold math

\begin{document}

\preprint{APS/123-QED}

\title{Consistency of shared reference frames should be reexamined}% Force line breaks with \\

\author{Fei Gao$^{1,2}$\footnote{Email: gaofei\_bupt@hotmail.com or hzpe@sohu.com}, \quad Fen-Zhuo Guo$^{1}$, \quad Qiao-Yan Wen$^{1}$, and Fu-Chen Zhu$^{3}$}
\affiliation{
$^{1}$State key Laboratory of Networking and Switching Technology, Beijing University of \\ Posts and Telecommunications, Beijing 100876, China\\
$^{2}$School of Science, Beijing University of Posts and
Telecommunications, Beijing 100876, China\\
$^{3}$National Laboratory for Modern Communications, P.O.Box 810,
Chengdu 610041, China}
%\date{\today}

\begin{abstract}
In a recent Letter [G. Chiribella et al., Phys. Rev. Lett.
\textbf{98}, 120501 (2007)], four protocols were proposed to
secretly transmit a reference frame. Here We point out that in
these protocols an eavesdropper can change the transmitted
reference frame without being detected, which means the
consistency of the shared reference frames should be reexamined.
The way to check the above consistency is discussed. It is shown
that this problem is quite different from that in previous
protocols of quantum cryptography.
\end{abstract}


\maketitle

As we know, reference frame (RF) \cite{BRS07} is a kind of
unspeakable information and consequently sharing of an RF is
generally more difficult than that of a string of key bits as
performed in a quantum key distribution (QKD)
protocol~\cite{GRTZ}. In a recent Letter, Chiribella et al.
proposed four quantum-cryptographic protocols to secretly
communicate an RF~\cite{CMP07}. These protocols are subtly
designed so that the eavesdropper (say Eve) cannot obtain any
information about the RF when it is transmitted between the users
(say Alice and Bob). Here, from a different perspective of
security, we consider a special threat which was not concerned in
Ref.~\cite{CMP07}. That is, after the communication, the
consistency of the RF Alice sent with that Bob received is still
not assured. We will show that by a special attack Eve can destroy
this consistency without introducing any detectable disturbance.

Let us take the first protocol in Ref.~\cite{CMP07} as our
example, where Alice can transmit a secret direction ($z$ axis) to
Bob with the help of a string of secret key bits shared between
them. Alice sends a sequence of spin-$\frac{1}{2}$ particles to
Bob, whose particular states, i.e. spin up or down, are determined
by the shared key bits 0 or 1, respectively. After receiving this
sequence, Bob measures the particles alternately along his own
$x$, $y$, and $z$ axes and compares the measurement results with
the shared key bits. By calculating the error rates associate to
the three measurement directions, Bob can estimate the angles
$\theta_x$, $\theta_y$, $\theta_z$ between his three axes $x$,
$y$, $z$ and Alice's $z$ axis [see Fig.1(a)]. Then Bob obtains the
direction of Alice's $z$ axis with a certain accuracy. To detect
eavesdropping, Bob checks whether the sum of the three angles'
squared cosines differs from 1 by more than an allowable error.

\begin{figure}
\includegraphics{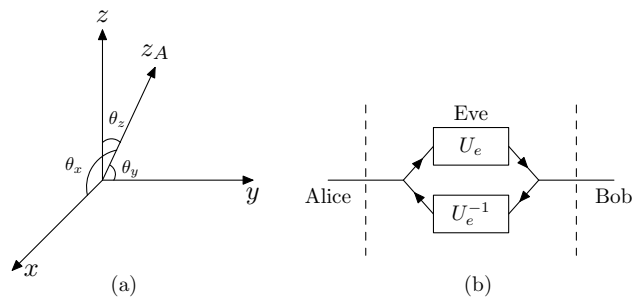}% Here is how to import EPS art
\caption{\label{fig:one} (a) The way to estimate the direction of
Alice's $z$ axis $z_A$. The three angles satisfy
$cos^2\theta_x+cos^2\theta_y+cos^2\theta_z=1$. (b) Eve's
additional strategy to avoid a general prepare-and-measure
detection.}
\end{figure}

Indeed, if Eve stays in the quantum channel and performs blindly
some measurements on the transmitted particles to extract
information of Alice's $z$ axis, her eavesdropping would result in
a depolarization of the spins and it follows that she would be
detected by Bob (the three angles recovered by Bob would be
inconsistent). However, Eve can take a special strategy to attack.
That is, she performs a unitary operation $U_e$ on each
transmitted particle to rotate the spins by a certain angle. Eve
does not know the value of this angle because she does not know
what on earth the operator $U_e$ is in Alice's representation
(therefore the choice of $U_e$ are at random). But she need not
know it. Here the purpose of Eve is to destroy the communication
instead of to extract information of the transmitted direction (In
fact, without the key bits shared between Alice and Bob, Eve can
never elicit any information about the transmitted direction by
measurements on these qubits, which is very similar to that in
one-time pad). Since the role of this attack is just rotate the
spins by a same angle, Bob will obtain a changed direction of
Alice's $z$ axis and, more seriously, he cannot detect the
presence of Eve (obviously for any direction, the three angles
$\eta_x$, $\eta_y$, $\eta_z$ between it and the three axes $x$,
$y$, $z$ satisfy $cos^2\eta_x+cos^2\eta_y+cos^2\eta_z=1$).

Similar risk exists in other three protocols in Ref.~\cite{CMP07},
where Alice would transmit a Cartesian frame to Bob secretly.
Consider any state $|\psi\rangle$ Alice will send to Bob. Since
$|\psi\rangle$ is generated by Alice according to her Cartesian
frame, from Bob's point of view all spins are rotated by a certain
rotation $g\in SU(2)$ which connects his axes with that of Alice.
Therefore, Bob's aim is to infer $g$ by measurements on all the
states Alice sent to him. This is the main idea of the three
protocols. Here Eve can also perform a unitary operation $U_e$ on
each particle (or a collective rotation $U^{\otimes N}_e$). Again,
Eve just chooses a unitary operation at random and does not know
the value of $e$ referring to Alice's Cartesian frame. Thus, after
some measurements, Bob's inferred parameter, which connects Alice
and Bob's axes, would be $eg$ but not $g$. Apparently, it looks
like that the parameter Alice sent is indeed $eg$ instead of $g$,
which is the only difference when this attack happens. So,
everything goes naturally. As a result, Bob obtains a Cartesian
frame different from Alice's and Eve would not be detected.

It can be seen that this special attack causes a severe effect,
that is, the RF cannot be successfully shared between Alice and
Bob as they intended. More seriously, in the end of the protocol,
when Alice and Bob are congratulated for the successful sharing of
the Cartesian frame, they even do not know they have been cheated
by Eve. When Alice and Bob utilize the different RFs to distribute
secret key bits~\cite{BRS04}, remarkable errors would appear. But
at that time they still do not know whether the RFs are
inconsistent or an eavesdropper exists in the process of key
distribution, which is a really intractable business.
Consequently, this problem must be overcome in a real
implementation. In fact, the above attack is a special kind of
denial-of-service attack, and it is also forbidden in a quantum
secure direct communication (QSDC) protocol~\cite{BF02,C03}.

To communicate something secretly, as generally required in
quantum cryptography, we have to take all thinkable attack
strategies into consideration. Otherwise the intended
communications may be attacked successfully (for instances, see
Refs.~\cite{C03,ZLG01,W04,GGW1,GGW2,DLZZ,GWZ07}). In
Ref.~\cite{CMP07} the main attention is paid to forbid Eve to
extract information of the transmitted RF, while the consistency
of the two RFs is overlooked.

Now consider how to detect this special attack. Different from a
general protocol of quantum cryptography, Alice and Bob have not a
shared RF before the communication. Therefore, the ability of
their possible ways to check eavesdropping is limited. For
example, as described in Ref.~\cite{CMP07}, they can only check
the uniqueness of the transmitted direction or frame, or employ
the rotationally invariant subsystems of the test qubits to
detect. But both of them are useless for the special attack.
Another immediate manner to detect is to compare the consistency
of Alice and Bob's RFs at the end of the communication. In a QKD
protocol Alice and Bob can sample some of the key bits to check if
the two keys are identical publicly. However, this strategy cannot
be used here because the transmitted object is unspeakable
information which cannot be discussed directly in a classical
channel.

An alternative way to detect the special attack is using the
general prepare-and-measure model. In particular, Alice (Bob)
prepares a certain state and sends it to Bob (Alice). After the
measurement of Bob (Alice), they can judge the consistency of
their RFs by comparing the initial state and the measurement
result in public. However, this strategy would not work if Eve
does an additional trick. That is, Eve performs the unitary
operation $U_e$ on all the qubits from Alice to Bob and another
one $U^{-1}_e$ on all the qubits from Bob to Alice [see Fig.1(b)].
As discussed in above paragraphs, the difference of Alice and
Bob's RFs is just $U_e$ (under the above attack). As a result,
under Eve's operations $U_e$ and $U^{-1}_e$, the prepared state by
Alice (Bob) and the measured state by Bob (Alice) would exhibit
the same physical features referring to their respective RFs.
Thus, no errors would appear when Alice and Bob do their
comparison.

In view of the above analysis, to expose the special attack, the
check qubits should not be transmitted in the communication
channel. A possible way is to use some states which are previously
shared between Alice and Bob. For example, Alice and Bob share
some entangled pairs in the spin singlet state
$|\Psi^-\rangle=(|01\rangle-|10\rangle)/\sqrt{2}$ before the
communication of RF (that is, for each pair, Alice holds one
photon and Bob controls the other). When a certain RF has been
transmitted Alice and Bob measure the spins of each shared singlet
state referring to their respective RFs (the RF Alice sent and the
one Bob received). Afterwards they compare the measurement results
publicly. If two RFs are identical the results should be
determinately anti-correlated. By this way, Alice and Bob can
assure the consistency of the two RFs (with certain accuracy) when
the error rate of this comparison is low enough. However, there is
a problem in this detection. That is, the requirement of shared
singlet pairs is too strong, with which Alice and Bob can even
achieve the whole transmission of a private RF \cite{fn1}.
Obviously, if Alice measures the spins of her photons in the
shared singlet pairs referring to her RF and announces the
results, Bob can obtain the RF by the technique similar to that in
the first protocol in Ref.~\cite{CMP07}.

One may argue that Eve can do the same sort of attack in many
other situations (for example, one-time pad, QKD, and QSDC) and
consequently it is not so meaningful to discuss this problem
further here. However, it is not the fact for the case of RF
sharing. As far as this attack is concerned, the transmission of
an RF and that of a classical or quantum message are quite
different. More specifically, when a message is transmitted, this
kind of attack can be easily avoided by the manners of message
authentication, error correction code, or directly sampling some
random transmitted bits/qubits to check eavesdropping \cite{C03}.
But all these manners are useless for the transmission of an RF.
From the above analysis we can see that this problem is
intractable and far from being totally resolved. It is urgent to
find an effective way to detect this attack (or check the
consistency of the shared RFs) in the transmission of an RF. As a
result, though this problem can be accepted as inevitable in
previous cryptographic models such as one-time pad, QKD, and QSDC,
it should be paid more attention in the case of RF sharing.

In conclusion, we present a special attack to the protocols of
secret communication of RF, by which an eavesdropper can change
the transmitted RF without being detected. It means that the
obtained RF should be reexamined in such protocols. Furthermore,
the way to check the consistency of the two RFs is discussed
though it needs more further study. Note that the communication of
RF is a new topic and it may interest many scholars. We hope that
the special attack is noticed and taken into account in the
following research.

We are grateful to the anonymous reviewer for helpful comments.
This work is supported by the National High Technology Research
and Development Program of China, Grant No. 2006AA01Z419; the
National Natural Science Foundation of China, Grant Nos. 90604023,
60373059; the National Research Foundation for the Doctoral
Program of Higher Education of China, Grant No. 20040013007; the
National Laboratory for Modern Communications Science Foundation
of China, Grant No. 9140C1101010601; the Natural Science
Foundation of Beijing, Grant No. 4072020; and the ISN Open
Foundation.

\end{document}